\documentclass[reqno]{amsart}

\usepackage{amsmath, amssymb}
\usepackage[mathscr]{eucal}
\usepackage{upgreek}
\usepackage{hyperref}
\hypersetup{colorlinks=true, linkcolor=blue, urlcolor=blue, citecolor=[rgb]{0, 0.8, 0}}

\theoremstyle{plain}
\newtheorem*{theorem}{Theorem}

\theoremstyle{remark}
\newtheorem{remark}{Remark}

\newcommand{\du}{\mathrm{d}}
\newcommand{\iu}{\mathrm{i}}

\DeclareMathOperator{\im}{Im}
\DeclareMathOperator{\ind}{ind}
\DeclareMathOperator{\Res}{Res}

\title{Spectral identities for Schr\"{o}dinger operators}

\author{Namig J. Guliyev}
\address{Institute of Mathematics and Mechanics, Azerbaijan National Academy of Sciences, 9 B.~Vahabzadeh str., AZ1141, Baku, Azerbaijan.}
\email{njguliyev@gmail.com}



\begin{document}
\maketitle
\begin{abstract}
  We obtain a system of identities relating boundary coefficients and spectral data for the one-dimensional Schr\"{o}dinger equation with boundary conditions containing rational Herglotz--Nevanlinna functions of the eigenvalue parameter. These identities can be thought of as a kind of mini version of the Gelfand--Levitan integral equation for boundary coefficients only.
\end{abstract}

\section{Introduction and main result} \label{sec:introduction}

In our recent papers, we solved various direct and inverse spectral problems for boundary value problems generated by the one-dimensional Schr\"{o}dinger equation
\begin{equation} \label{eq:SL}
  -y''(x) + q(x)y(x) = \lambda y(x)
\end{equation}
and the boundary conditions
\begin{equation} \label{eq:boundary}
  \frac{y'(0)}{y(0)} = -f(\lambda), \qquad \frac{y'(\pi)}{y(\pi)} = F(\lambda),
\end{equation}
where $f$ and $F$ are rational Herglotz--Nevanlinna functions. As the papers~\cite{G17} and \cite{G19a} show, for summable and distributional potentials respectively, their spectral theory is mostly analogous to the classical case when the coefficients $f$ and $F$ in~(\ref{eq:boundary}) are constants. There also appear some new phenomena. For example, two spectra may no longer be sufficient to determine the coefficients of~(\ref{eq:SL})-(\ref{eq:boundary}) uniquely~\cite{G18} and it is quite possible for eigenfunctions not to be a basis for $\mathscr{L}_2(0,\pi)$, even after removing some of them~\cite{G19b}. However, it is still possible to determine the coefficients of~(\ref{eq:SL})-(\ref{eq:boundary}) from the knowledge of its spectral data consisting of the eigenvalues and the so-called norming constants (see~(\ref{eq:gamma}) below). The purpose of this short paper is to establish some simple identities relating the boundary coefficients (i.e., $f$ and $F$) and the spectral data of~(\ref{eq:SL})-(\ref{eq:boundary}) with a real-valued potential $q \in \mathscr{L}_1(0, \pi)$, and which, in particular, allow one to find the boundary coefficients in a much more direct way. Remarkably, some of the objects that we use in this paper, namely the polynomial $\boldsymbol{\upomega}_f$ and the matrices of~(\ref{eq:identities}), are very similar to, but not quite identical with, those encountered in other areas of mathematics~\cite{HT12}.

Apart from being interesting in their own right, eigenvalue problems with boundary conditions dependent on the eigenvalue parameter have found numerous applications in various fields of science. For example, they have recently been considered in the literature in connection with such diverse areas as heat transfer~\cite{BHM15}, string theory~\cite{GKR16}, fluid dynamics~\cite{KG15}, biology~\cite{KM19}, mathematical finance~\cite{NL19}, and quantum computing~\cite{PRSE18}. Interestingly, the last-mentioned paper also discusses some identities of the kind considered in Remark~\ref{rem:identities} below.

We now recall some necessary notation from~\cite{G17} related to rational Herglotz--Nevanlinna functions. Each such function can be written as
\begin{equation} \label{eq:f_F}
  f(\lambda) = h_0 \lambda + h + \sum_{k=1}^d \frac{\delta_k}{h_k - \lambda}
\end{equation}
with $h_0 \ge 0$, $h \in \mathbb{R}$, $\delta_k > 0$, and $h_1 < \ldots < h_d$. To every function $f$ of this form we assign two polynomials $f_\uparrow$ and $f_\downarrow$ by writing this function as
$$
  f(\lambda) = \frac{f_\uparrow(\lambda)}{f_\downarrow(\lambda)},
$$
where
$$
  f_\downarrow(\lambda) := h'_0 \prod_{k=1}^d (h_k - \lambda), \qquad h'_0 := \begin{cases} 1 / h_0, & h_0 > 0, \\ 1, & h_0 = 0. \end{cases}
$$
We also associate to $f$ its \emph{index}
$$
  \ind f := \begin{cases} 2 d + 1, & h_0 > 0, \\ 2 d, & h_0 = 0, \end{cases}
$$
which counts each finite pole of this function twice and its pole at infinity (if any) once. To each rational Herglotz--Nevanlinna function $f$ we also assigned in \cite{G17} a monic polynomial
\begin{equation*}
  \boldsymbol{\upomega}_f(\lambda) := (-1)^{\left\lfloor \frac{\ind f}{2} \right\rfloor} \lambda f_\downarrow \left( \lambda^2 \right) - (-1)^{\left\lceil \frac{\ind f}{2} \right\rceil} f_\uparrow \left( \lambda^2 \right),
\end{equation*}
where $\lfloor \cdot \rfloor$ and $\lceil \cdot \rceil$ are the usual floor and ceiling functions.

Now we are ready to introduce the main players of this paper. Denote by $\lambda_n$ the eigenvalues of the boundary value problem~(\ref{eq:SL})-(\ref{eq:boundary}) and by $\gamma_n$ its norming constants defined by~(\ref{eq:gamma}) below. Consider the sums of the series
\begin{equation*}
  \sigma_k := \sum_{n=0}^\infty \frac{\lambda_n^k}{\gamma_n}, \qquad k = 0, 1, \ldots, \ind f - 1
\end{equation*}
and
\begin{equation*}
  \sigma_{\ind f} := \sum_{n < L} \frac{\lambda_n^{\ind f}}{\gamma_n} + \sum_{n = L} \left( \frac{\lambda_n^{\ind f}}{\gamma_n} - \frac{1}{\pi} \right) + \sum_{n > L} \left( \frac{\lambda_n^{\ind f}}{\gamma_n} - \frac{2}{\pi} \right)
\end{equation*}
with
\begin{equation*}
  L := \frac{\ind f + \ind F}{2},
\end{equation*}
whose convergence is an immediate consequence of the asymptotics of the eigenvalues and the norming constants \cite[Theorem 4.2]{G17}. Alternatively, their convergence also follows from the proof below (see Section~\ref{sec:proof}). On the other hand, the above polynomial $\boldsymbol{\upomega}_f$ can be written as
\begin{equation*}
  \boldsymbol{\upomega}_f(\lambda) = \lambda^{\ind f + 1} + \omega_1 \lambda^{\ind f} + \ldots + \omega_{\ind f + 1}.
\end{equation*}
It turns out that these two $(\ind f + 1)$-tuples $\sigma_0$, $\ldots$, $\sigma_{\ind f}$ and $\omega_1$, $\ldots$, $\omega_{\ind f + 1}$ are related in a rather simple way.

\begin{theorem}
  The following identities hold:
\begin{equation} \label{eq:identities}
  (-1)^k \omega_{1 - k} + \sum_{i = - \lfloor k / 2 \rfloor}^{\lfloor (\ind f + 1 - k) / 2 \rfloor} \sigma_{\ind f - i - k} \omega_{2 i + k} = 0, \qquad k = 0, 1, \ldots, \ind f
\end{equation}
with the convention that $\omega_0 = 1$ and $\omega_m = 0$ if $m < 0$. Moreover, this system is uniquely solvable, both when either $\omega_1$, $\ldots$, $\omega_{\ind f + 1}$ or $\sigma_0$, $\ldots$, $\sigma_{\ind f}$ are treated as variables.
\end{theorem}

We will prove this theorem in the next section. It is interesting to note that the system~(\ref{eq:identities}) resembles, in a sense, the Gelfand--Levitan integral equation. Indeed, the numbers $\omega_k$ are defined in terms of the coefficient $f$ and play the role of the Gelfand--Levitan kernel. The numbers $\sigma_k$, on the other hand, are defined in terms of the eigenvalues and the norming constants, and thus play the role of the other function of two variables from the Gelfand--Levitan equation, usually denoted by $f(\cdot, \cdot)$ or $F(\cdot, \cdot)$.

\begin{remark} \label{rem:identities}
  The above theorem allows one to determine $f$ in terms of the coefficients of the polynomials $f_\downarrow$ and $f_\uparrow$. It is also possible to obtain some identities involving the coefficients in the representation~(\ref{eq:f_F}). Recall that \cite[Section 2.2]{G17} the boundary value problem~(\ref{eq:SL})-(\ref{eq:boundary}) can also be stated as an eigenvalue problem for a self-adjoint operator in a Hilbert space of the form $\mathscr{L}_2 \oplus \mathbb{C}^M$ for a suitable $M$, namely
\begin{equation*}
  M := \left\lceil \frac{\ind f}{2} \right\rceil + \left\lceil \frac{\ind F}{2} \right\rceil.
\end{equation*}
Parseval's identity for the eigenvectors of this operator yields the identities
\begin{equation*}
  \frac{1}{\delta_k} = \sum_{n=0}^\infty \frac{1}{\gamma_n} \left( \frac{f_\downarrow(\lambda_n)}{\lambda_n - h_k} \right)^2, \qquad k = 1, \ldots, d
\end{equation*}
and
\begin{equation*}
  \frac{1}{h_0} = \sum_{n=0}^\infty \frac{f_\downarrow^2(\lambda_n)}{\gamma_n},
\end{equation*}
including the possibility that both sides of this last identity are infinite. However, it seems that we still need the above theorem to determine $h_1$, $\ldots$, $h_d$, or equivalently $f_\downarrow$. Such identities were obtained in~\cite{G06} for the case when $\ind f = \ind F = 2$.
\end{remark}

\begin{remark} \label{rem:beta}
Of course, all of the above identities have their counterparts for the right endpoint. One only needs to replace $\gamma_n$ by $\beta_n^2 \gamma_n$ everywhere, where $\beta_n$ is defined by~(\ref{eq:beta}) below. This can be proven by repeating the corresponding arguments for the left endpoint. A much shorter proof can also be given, based on a simple symmetry argument: for the boundary value problem of the form~(\ref{eq:SL})-(\ref{eq:boundary}) with $q(x)$, $f$, and $F$ replaced by $q(\pi-x)$, $F$, and $f$ respectively, the roles of the solutions $\varphi$ and $\psi$ defined by~(\ref{eq:phi_psi}) are interchanged, and hence that problem has the same eigenvalues $\lambda_n$ and the norming constants $\beta_n^2 \gamma_n$.

\end{remark}

In addition to being interesting for their own sake, these identities also have a multitude of applications to inverse problems. For example, as we have already pointed out, they allow one to avoid the whole machinery of \cite{G17} and compute the coefficients of $f$ directly. Moreover, for constant $f$ (i.e., when $\ind f = 0$), the system~(\ref{eq:identities}) consists of only one identity and this identity is a key ingredient in Korotyaev and Chelkak's parametrization of sets of isospectral problems~\cite{KC09}. The identities~(\ref{eq:identities}) also play an important role in the solution of the two-spectra inverse problem~\cite{G18} for boundary value problems of the form~(\ref{eq:SL})-(\ref{eq:boundary}).

The result of this paper can also be useful in the so-called incomplete inverse problems. To determine the boundary value problem~(\ref{eq:SL})-(\ref{eq:boundary}) completely, one needs to know all of the eigenvalues and the norming constants. But if we have some extra information about the coefficients of this boundary value problem, a part of this spectral data can be omitted (see, e.g., \cite[Section 4.6]{G17}, \cite{WW18}, and the references therein). In particular, if the boundary coefficients $f$ and $F$ are known and some of the eigenvalues and the norming constants are missing, with the number of missing ones not exceeding $\ind f + \ind F + 2$, then the identities~(\ref{eq:identities}) (and their counterparts for the right endpoint, mentioned in Remark~\ref{rem:beta}) give us algebraic equations for these missing ones.

\section{Proof} \label{sec:proof}

Before turning to the proof, we need some more definitions and preliminary results. Let $\varphi(x, \lambda)$ and $\psi(x, \lambda)$ be the solutions of (\ref{eq:SL}) satisfying the initial conditions
\begin{equation} \label{eq:phi_psi}
  \varphi(0, \lambda) = f_\downarrow(\lambda), \quad \varphi'(0, \lambda) = -f_\uparrow(\lambda), \quad \psi(\pi, \lambda) = F_\downarrow(\lambda), \quad \psi'(\pi, \lambda) = F_\uparrow(\lambda).
\end{equation}
Then the eigenvalues of the boundary value problem (\ref{eq:SL})-(\ref{eq:boundary}) coincide with the zeros of the \emph{characteristic function}
\begin{equation*}
  \chi(\lambda) := F_\uparrow(\lambda) \varphi(\pi, \lambda) - F_\downarrow(\lambda) \varphi'(\pi, \lambda) = f_\downarrow(\lambda) \psi'(0, \lambda) + f_\uparrow(\lambda) \psi(0, \lambda)
\end{equation*}
and for each eigenvalue $\lambda_n$ there exists a unique number $\beta_n \ne 0$ such that
\begin{equation} \label{eq:beta}
  \psi(x, \lambda_n) = \beta_n \varphi(x, \lambda_n).
\end{equation}
We define the \emph{norming constants} as
\begin{equation} \label{eq:gamma}
  \gamma_n := \int_0^{\pi} \varphi^2(x, \lambda_n) \,\du x + f'(\lambda_n) \varphi^2(0, \lambda_n) + F'(\lambda_n) \varphi^2(\pi, \lambda_n),
\end{equation}
where the second (respectively, the third) summand on the right-hand side is omitted if $\lambda_n$ coincides with one of the poles of the function $f$ (respectively, $F$). The three sequences $\{ \lambda_n \}_{n \ge 0}$, $\{ \beta_n \}_{n \ge 0}$, and $\{ \gamma_n \}_{n \ge 0}$ are related by the identity (\cite[Lemma~2.1]{G17})
\begin{equation} \label{eq:chi_beta_gamma}
  \chi'(\lambda_n) = \beta_n \gamma_n.
\end{equation}

The solution $\psi$ and its first derivative with respect to $x$ satisfy the asymptotic estimates
\begin{multline*}
  \psi(0, \lambda) = \lambda^{\ind F / 2} \cos \left( \sqrt{\lambda} + \frac{\ind F}{2} \right) \pi \\
  + \left( \frac{1}{2} \int_0^\pi q(x) \,\du x + \Omega_1 \right) \lambda^{(\ind F - 1) / 2} \sin \left( \sqrt{\lambda} + \frac{\ind F}{2} \right) \pi \\
  + o\left( \lambda^{(\ind F - 1) / 2} e^{|\im \sqrt{\lambda}\pi|} \right)
\end{multline*}
and
\begin{multline*}
  \psi'(0, \lambda) = \lambda^{(\ind F + 1) / 2} \sin \left( \sqrt{\lambda} + \frac{\ind F}{2} \right) \pi \\
  - \left( \frac{1}{2} \int_0^\pi q(x) \,\du x + \Omega_1 \right) \lambda^{\ind F / 2} \cos \left( \sqrt{\lambda} + \frac{\ind F}{2} \right) \pi \\
  + o\left( \lambda^{\ind F / 2} e^{|\im \sqrt{\lambda}\pi|} \right),
\end{multline*}
which in turn imply the asymptotic formula
\begin{multline*}
  \chi(\lambda) = \lambda^{L + 1/2} \sin \left( \sqrt{\lambda} + L \right) \pi \\
  - \left( \frac{1}{2} \int_0^\pi q(x) \,\du x + \omega_1 + \Omega_1 \right) \lambda^L \cos \left( \sqrt{\lambda} + L \right) \pi + o\left( \lambda^L e^{|\im \sqrt{\lambda}\pi|} \right)
\end{multline*}
for the characteristic function. These estimates can be obtained by expressing $\psi$ as a linear combination of the cosine- and sine-type solutions and using the well-known estimates for them. Another possible way to obtain these formulas is to observe that they all are (up to a change in sign) the characteristic functions of problems of the form~(\ref{eq:SL})-(\ref{eq:boundary}) (with the Dirichlet condition at the left endpoint in the case of $\psi(0, \lambda)$ and the Neumann condition in the case of $\psi'(0, \lambda)$) and then apply a lemma of Marchenko and Ostrovskii (see \cite[Lemma~3.4.2]{M77}) to their infinite product representations (cf. \cite[Appendix~A]{G18}).

We are now going to obtain the identities~(\ref{eq:identities}) as a consequence of a well-known result from complex analysis: if $g$ is a meromorphic function satisfying the estimate
\begin{equation*}
  \sup_{|\lambda| = R_N} |g(\lambda)| = o\left( \frac{1}{R_N} \right)
\end{equation*}
with
\begin{equation*}
  R_N := \left( N - L - \frac{1}{2} \right)^2,
\end{equation*}
then the sum of the residues of this function, defined as $\frac{1}{2 \pi \iu} \lim_{N \to \infty} \int_{|\lambda| = R_N} g(\lambda) \,\du \lambda$, is zero \cite[Lemma 3.2]{PT87}.

It is convenient to consider the cases of odd and even $\ind f$ separately. However, since the arguments are very similar in both cases, we give the details only for the former case. Taking $g(\lambda) = \lambda^j \psi(0, \lambda) / \chi(\lambda)$ and $g(\lambda) = \lambda^j \psi'(0, \lambda) / \chi(\lambda)$ respectively in the above result for $j = 0, 1, \ldots, d-1$ and using~(\ref{eq:chi_beta_gamma}), we obtain
\begin{equation*}
  \sum_{n=0}^\infty \frac{\lambda_n^j f_\downarrow(\lambda_n)}{\gamma_n} = \sum_{n=0}^\infty \frac{\lambda_n^j \beta_n f_\downarrow(\lambda_n)}{\chi'(\lambda_n)} = \sum_{n=0}^\infty \Res_{\lambda = \lambda_n} \frac{\lambda^j \psi(0, \lambda)}{\chi(\lambda)} = 0
\end{equation*}
and
\begin{equation*}
  \sum_{n=0}^\infty \frac{\lambda_n^j f_\uparrow(\lambda_n)}{\gamma_n} = \sum_{n=0}^\infty \frac{\lambda_n^j \beta_n f_\uparrow(\lambda_n)}{\chi'(\lambda_n)} = - \sum_{n=0}^\infty \Res_{\lambda = \lambda_n} \frac{\lambda^j \psi'(0, \lambda)}{\chi(\lambda)} = 0,
\end{equation*}
which are exactly the identities~(\ref{eq:identities}) for $k = 2, 3, \ldots, \ind f$. To obtain the case $k = 1$ we take
\begin{equation*}
  g(\lambda) = \frac{\lambda^d \psi(0, \lambda)}{\chi(\lambda)} - \frac{(-1)^d}{\lambda}.
\end{equation*}
Finally, the case $k = 0$ is obtained by taking
\begin{multline*}
  g(\lambda) = \frac{(-\lambda)^d \psi'(0, \lambda)}{\chi(\lambda)} + \lambda^{-1/2} \cot \left( \sqrt{\lambda} + L \right) \pi \\
  + \left( \frac{1}{2} \int_0^\pi q(x) \,\du x + \omega_1 + \Omega_1 \right) \lambda^{-1} \cot^2 \left( \sqrt{\lambda} + L \right) \pi \\
  + \left( \frac{1}{2} \int_0^\pi q(x) \,\du x + \Omega_1 \right) \lambda^{-1}
\end{multline*}
and using the above asymptotic formulas for $\psi$, $\psi'$, and $\chi$ together with the obvious equalities
\begin{equation*}
  \Res_{\lambda = (n-L)^2} \lambda^{-1/2} \cot \left( \sqrt{\lambda} + L \right) \pi = \begin{cases} \frac{2}{\pi}, & n > L, \\ \frac{1}{\pi}, & n = L \end{cases}
\end{equation*}
and
\begin{equation*}
  \Res_{\lambda = (n-L)^2} \lambda^{-1} \cot^2 \left( \sqrt{\lambda} + L \right) \pi = \begin{cases} - \frac{2}{\pi^2 (n-L)^2}, & n > L, \\ - \frac{2}{3}, & n = L. \end{cases}
\end{equation*}

We now turn to uniqueness. When $\omega$s are treated as variables, taking into account the fact that all the equations of the system~(\ref{eq:identities}) except the one with $k = 0$ contain only $\omega$s with indices of the same parity, this system can be solved in two steps: one first determines $\omega_k$ with $k$ of the same parity as $\ind f$ (i.e., those coming from $f_\downarrow$) and then finds the remaining ones. More precisely, if $\ind f = 2 d + 1$ then the subsystem of~(\ref{eq:identities}) consisting of the equations with odd $k$ is a system of linear equations with respect to $\omega_1$, $\omega_3$, $\ldots$, $\omega_{2 d + 1}$ whose matrix is a strictly positive definite Hankel matrix, and is thus uniquely solvable (see the proof of~\cite[Lemma 2.1]{G18} for details). Now it only remains to observe that the matrix of the subsystem of~(\ref{eq:identities}) with even $k$ is the same strictly positive definite Hankel matrix as above. Similarly, if $\ind f = 2 d$ then $\omega_2$, $\omega_4$, $\ldots$, $\omega_{2 d}$ are uniquely determined from the subsystem of~(\ref{eq:identities}) consisting of the equations with nonzero even $k$, $\omega_1$ is then found from the equation with $k = 0$, and finally the remaining $\omega$s are determined from the equations with odd $k$.

Finally, when $\sigma$s are treated as variables, the determinant of the system~(\ref{eq:identities}) is simply equal to the resultant of the polynomials $f_\downarrow$ and $f_\uparrow$, up to a possible change in sign. As these two polynomials have no common roots, their resultant is nonzero.

\end{document}